\documentclass[a4paper,pra,showpacs,superscriptaddress,twocolumn]{revtex4-2}

\usepackage[utf8]{inputenc}
\usepackage[T1]{fontenc}
\usepackage{amsfonts}
\usepackage{mathrsfs}
\usepackage{amsmath}
\usepackage{amssymb}
\usepackage{amsthm}
\usepackage{graphicx}
\usepackage{braket}
\usepackage{hyperref}
\hypersetup{colorlinks=true,linkcolor=blue,anchorcolor=blue,citecolor=blue}
\usepackage{bbm,bm}

\newcommand{\myeq}[1]{Eq.\hspace{0.3em}\eqref{#1}}

\newcommand{\mycite}[1]{Ref.\hspace{0.3em}\cite{#1}}
\newcommand{\mycites}[1]{Refs.\hspace{0.3em}\cite{#1}}
\newcommand{\myfig}[1]{Fig.\hspace{0.3em}\ref{#1}}
\newcommand{\myapp}[1]{Appendix.\hspace{0.3em}\ref{#1}}

\newcommand{\mysec}[1]{Sec.\hspace{0.3em}\ref{#1}}
\begin{document}

\title{Entropic uncertainty relations and entanglement detection from quantum designs}

\author{Yundu Zhao}
\thanks{These two authors contributed equally to this work.}
\affiliation{National Laboratory of Solid State Microstructures and School of Physics, Collaborative Innovation Center of Advanced Microstructures, Nanjing University, Nanjing 210093, China}
\affiliation{Institute for Brain Sciences and Kuang Yaming Honors School, Nanjing University, Nanjing 210023, China}
\affiliation{Hefei National Laboratory, University of Science and Technology of China, Hefei 230088, China}

\author{Shan Huang}
\thanks{These two authors contributed equally to this work.}
\affiliation{National Laboratory of Solid State Microstructures and School of Physics, Collaborative Innovation Center of Advanced Microstructures, Nanjing University, Nanjing 210093, China}
\affiliation{Institute for Brain Sciences and Kuang Yaming Honors School, Nanjing University, Nanjing 210023, China}
\affiliation{Hefei National Laboratory, University of Science and Technology of China, Hefei 230088, China}

\author{Shengjun Wu}
\email{sjwu@nju.edu.cn} 
\affiliation{National Laboratory of Solid State Microstructures and School of Physics, Collaborative Innovation Center of Advanced Microstructures, Nanjing University, Nanjing 210093, China}
\affiliation{Institute for Brain Sciences and Kuang Yaming Honors School, Nanjing University, Nanjing 210023, China}
\affiliation{Hefei National Laboratory, University of Science and Technology of China, Hefei 230088, China}

\date{\today}
	
\begin{abstract}
Uncertainty relations and quantum entanglement are pivotal concepts in quantum theory.
Beyond their fundamental significance in shaping our understanding of the quantum world, they also underpin crucial applications in quantum information theory. 
In this article, we investigate entropic uncertainty relations and entanglement detection with an emphasis on quantum measurements with design structures. On the one hand, we derive improved Rényi entropic uncertainty relations for design-structured measurements, exploiting the property that the sum of powered (e.g., squared) probabilities of obtaining different measurement outcomes is now invariant under unitary transformations of the measured system and can be easily computed. On the other hand, the above property essentially imposes a state-independent upper bound, which is achieved at all pure states, on one's ability to predict local outcomes when performing a set of design-structured measurements on quantum systems. Realizing this, we also obtain criteria for detecting multi-partite entanglement with design-structured measurements.
\end{abstract}

\maketitle

\section{INTRODUCTION}\label{sec_1}
Quantum measurements are fundamental elements of quantum theory, which are essential tools for extracting the intrinsic information contained in quantum systems.
With the development of quantum information theory, measurements have also been recognized as important resources for quantum computation and communication \cite{resource_1,resource_2}. 
In particular, (an incomplete subset of) design-structured measurements (DSMs) such as measurements in mutually unbiased bases (MUBs) \cite{MUB_1, MUB_2, MUB_3, MUB_4}, mutually unbiased measurements \cite{MUM_1, MUM_2, MUM_3} and symmetric informationally complete measurements \cite{SIC_1, SIC_2, SIC_3},  are frequently utilized in various quantum information processing tasks, including entanglement detection \cite{ED_source_and_quantity, ED_use, huang_new, ED_SIC}, quantum state estimation \cite{MUB_tom_1, MUB_tom_2, SIC_tom}, and quantum key distribution \cite{MUB_QKD_1, MUB_QKD_2, SIC_QKD}, due to the practical significance or advantages \cite{MUB_tom_1, MUB_tom_2} endowed by their symmetric structures.

To harness general DSMs in sophisticated quantum applications, it is essential to see how DSMs can be employed to demonstrate fundamental quantum concepts like Heisenberg's uncertainty principle \cite{heisenberg}.  
Among all kinds of uncertainty measures, entropies receive special attention for their  operational significance from an information-theoretical perspective \cite{EUR_1, EUR_2, EUR_notable}. 
For example, Maassen and Uffink \cite{EUR_notable} proposed the famous entropic uncertainty relation (EUR) for two non-degenerate observables $A$ and $B$ of $d$-dimensional quantum systems
\begin{equation}
H(A)+H(B)\geq-\ln c_{\rm max},\label{notable_EUR}
\end{equation}
where $H=-\sum_ip_i\ln p_i$ refers to the Shannon entropy \cite{shannon} of probability distribution induced by measuring observables $A$ or $B$, and $c_{\rm max}=\max_{i,j}\{|\braket{a_i|b_j}|^2\}$ denotes the maximal overlap between eigenbases $\{\ket{a_i}\}$ of $A$ and $\{\ket{b_i}\}$ of  $B$. 
Notably, \myeq{notable_EUR} is tight when observables $A$ and $B$ are complementary, i.e.,  $\{\ket{a_i}\}$ and $\{\ket{b_i}\}$ are mutually unbiased in the sense that $|\braket{a_i|b_j}|^2=1/d$ for all $1\leq i,j\leq d$. In this case, an equality of \myeq{notable_EUR} can be achieved by quantum systems in any eigenstate of observable $A$  ($B$), which means the outcomes of measuring observable $B$ $(A)$ would be completely random with the corresponding entropy maximized. In contrast, no uncertainty exists for compatible observables since $c_{\rm max}=1$, and any quantum system in the common eigenstate of observables $A$ and $B$ leads to $H(A)=H(B)=0$.

EURs play an essential role in diverse quantum tasks \cite{EUR_review_1, EUR_review_2}, e.g., in the security analysis of quantum cryptography \cite{ quantum_cryptography_review, quantum_cryptography_1, quantum_cryptography_2}. A great deal of effort has been devoted to establishing EURs involving multiple measurements \cite{ref_43,multiple_1, multiple_5, multiple_6, multiple_7, multiple_3, wu, MUB_SIC, ref_48, huang_old, ref_42, huang_new_, ket, ras} since  Maassen and Uffink \cite{EUR_notable}.
When considering multiple measurements, however, it is unsurprising that EURs in terms of a multi-bases generalization \cite{multiple_6} of the maximal overlap between measurements are not always strong \cite{huang_new_}, since simply the maximal overlap could be too rough a characterization of measurement structure which underlies the uncertainty. On the other hand, it turns out that EURs \cite{multiple_3, wu, MUB_SIC, ref_48, huang_old, ref_42, huang_new_, ket, ras}, formulated from the indexes of coincidence (IC) of probabilities distributions of measurement outcomes, can be relatively strong, especially for a sufficiently large number of measurement bases that are close to being mutually unbiased \cite{huang_new_}.

The IC of some probability distribution $P=(p_1,p_2,\cdots)$ refers to the probability, denoted $I(P)$, that two random events drawn from $P$ are identical,  i.e., $I(P)=\sum_i p_i^2$.  As a direct generalization of the conventional IC,  $I_a(P)=\sum_ip_i^a$   ($a\geq2$ is an arbitrary integer that is equal to or smaller than some integer $t$ associated with the specific DSMs under consideration) will be called the $a$th order IC. 
Essentially, the IC is a measure of certainty. Consider a number $\Theta$ of measurements $\{P_{\theta}\}=\{p_{i|\theta}\}$, with $p_{i|\theta}$ denoting the probability of obtaining the $i$th outcome when performing the $\theta$th measurement. 
Obviously, no uncertainty exists when and only when $\sum_{\theta}I_a(P_{\theta})=\Theta$, in which case the outcomes of all the $\Theta$ measurements can be predicted with certainty, i.e., $\max_i\{ p_{i|\theta}\}=1$ for any $\theta=1,\cdots,\Theta$.  
Previous EURs for DSMs \cite{multiple_3, wu, MUB_SIC, ref_48, huang_old, ref_42, huang_new_, ket, ras} mainly utilize the property of outcome probability distributions induced by performing DSMs on quantum systems below 
\begin{equation}\label{IC_def}
\frac{1}{\Theta}\sum_{\theta}I_a(P_{\theta})=\frac{1}{\Theta}\sum_{i,\theta}p_{i|\theta}^{a}=B_a(\rho),
\end{equation}
where $B_a(\rho)$ can be easily computed from the DSMs and the state $\rho$ of the measured system \big(see \myeq{design_IC}\big).

EURs for DSMs have already been largely investigated in the special case $a=2$ \cite{multiple_3, wu, MUB_SIC, ref_48, huang_old, ref_42, huang_new_}, whereas for general DSMs, as we will see later, the results obtained in the previous work \cite{ket, ras} can be considerably tightened.  
In this work, we establish EURs for DSMs in terms of generalized entropies \cite{renyi} based on the corresponding IC \eqref{IC_def}. 
Furthermore, we demonstrate that our EURs are improved than the previous ones \cite{ket, ras} for general DSMs, which would consequently be advantageous in practical applications.
On the way, we also explore the applications of DSMs in detecting multi-partite entanglement.

This paper is structured as follows. We revisit some basic properties about general DSMs in \mysec{sec_2},  based on which we proceed to formulate improved EURs for DSMs in  \mysec{sec_3}.  In \mysec{sec_4}, we investigate the applications of DSMs in entanglement detection. In \mysec{sec_5},  we have some further discussions and draw a brief conclusion.

\section{DESIGN-STRUCTURED MEASUREMENTS}\label{sec_2}

The general quantum measurement is described by positive-operator-valued measures (POVMs) $\mathcal{M}$, which consists of positive semi-definite operators (called effects) $M_i\geq 0$, that sum up to the identity operator $\mathbbm{1}$,  $\sum_{i}M_{i}=\mathbbm{1}$. 
When applying the POVM $\mathcal{M}=\{M_i\}$ to a quantum system in the state $\rho$, according to Born's rule, the outcome probabilities are  $p_i=\mathrm{tr}(M_i\rho)$.
With $M_{i|{\theta}}$ denoting the $i$th effect of the $\theta$th POVM $\mathcal{M}_{\theta}$, we say a set of POVMs $\{\mathcal{M}_{\theta}\}=\{{M}_{i|\theta}\}$ is design-structured if the $t$-fold $(t\geq2)$ product of effects $\{M_{i|\theta}^{\otimes t}\}$, defined on the $t$-fold product space $\mathcal{H}_{d}^{\otimes t}$, sum up to the identity $\mathbbm{1}_{d^{t}}^{\rm sym}$ on the symmetric subspace of $\mathcal{H}_{d}^{\otimes t}$ up to a constant factor. Typical DSMs include arbitrary measurements in $d+1$ MUBs of $d$-dimensional Hilbert space $\mathcal{H}_d$.

Our main focus will be on DSMs that can be constructed from quantum $t$-designs \cite{ket}. A set of unit vectors $\{\ket{\psi_k}\}_{k=1}^K$ of $\mathcal{H}_{d}$ is said to be a quantum  $t$-design if, for an arbitrary polynomial of degree at most $t$, denoted as $P_t$, it holds that
\begin{align}
\frac{1}{K}\sum_{k=1}^KP_t(\ket{\psi_k})=\int d\psi P_t(\ket{\psi_{k}}),
\label{design}
\end{align}
where the integration is over the Harr measure \cite{harr} on the unit sphere of $\mathcal{H}_d$. Note here this definition \eqref{design} implies that a quantum $t$-design must also be an $a$-design whenever $a\leq t$.

To construct DSMs, we will frequently utilize the following property of quantum $t$-design: up to a constant factor,  the rank-1 projectors $\{|\psi_{k}\rangle\langle\psi_{k}|^{\otimes t}\}_{k=1}^K$ on $\mathcal{H}^{\otimes t}$ sum up to the identity operator $\mathbbm{1}_{d^{t}}^{\rm sym}$ on the symmetric subspace of $\mathcal{H}^{\otimes t}$, or formally, 
\begin{equation}
\sum_{k=1}^{K}\ket{\psi_{k}}\bra{\psi_{k}}^{\otimes t}=K\mathcal{D}_{d}^{(t)}\mathbbm{1}_{d^{t}}^{\rm sym},
\label{design_pro}
\end{equation}
where  $\mathcal{D}_{d}^{(t)}=\frac{t!(d-1)!}{(t+d-1)!}=1/tr(\mathbbm{1}_{d^{t}}^{\rm sym})$ . 

It is worth mentioning that those POVMs whose effects are proportional to the above rank-1 projectors $\{\ket{\psi_{k}}\bra{\psi_{k}}\}_{k=1}^K$ are necessarily DSMs. 
To be specific, let $\{\ket{\psi_k}\}_{k=1}^K$ be a quantum $t$-design, then the POVM  $\mathcal{M}=\{\frac{d}{K}\ket{\psi_{k}}\bra{\psi_{k}}\}_{k=1}^K$ is a DSM.  
Furthermore, if the rank-1 projectors from a $t$-design can be classified into $\Theta$ groups  $\{\{|\psi_{i|\theta}\rangle\langle\psi_{i|\theta}|\}_{i=1}^{K/\Theta}\}_{\theta=1}^{\Theta}$ such that the sum of  elements of each group  is proportional to the identity $\mathbbm{1}_d$, then $\Theta$ POVMs can be constructed from this quantum design as follows
\begin{equation}\label{design_con}
M_{i|{\theta}}=\frac{d}{L}\ket{\psi_{i|{\theta}}}\bra{\psi_{i|{\theta}}},
\end{equation}
where $L=K/\Theta$ denotes the number of POVM effects contained in individual measurements. 

Now observe from \myeq{design_pro} that projecting a $t$-fold product of density operator $\rho^{\otimes t}$ onto the symmetric subspace leads us to the equality
\begin{equation}\label{design_pro_2}
\sum_{k=1}^{K}\bra{\psi_{k}}\rho\ket{\psi_{k}}^{t}=K\mathcal{D}_{d}^{(t)}F_{t}(\rho),
\end{equation} 
where $\frac{(t+d-1)!}{t!(d-1)!d^t}\leq F_{t}(\rho)=\mathrm{tr}(\mathbbm{1}_{d^{t}}^{\rm sym}\rho^{\otimes t})\leq 1$.  
Interestingly, as has been noted in \mycite{ket}, the quantity $F_{t}(\rho)$ can be written as a sum of monomials of  $tr(\rho^{a})$, e.g., $ F_{2}(\rho)=\frac{1}{2}\big(1+\mathrm{tr}(\rho^{2})\big)$, $F_{3}(\rho)=\frac{1}{6}\big(1+3\mathrm{tr}(\rho^{2})+2\mathrm{tr}(\rho^{3})\big)$.
Moreover, combining Eqs.\hspace{0.3em}(\ref{design_con},\ref{design_pro_2}) we immediately have the average $a$th order IC of probability distributions induced by performing DSMs constructed from a quantum $t$-design 
\begin{align}\label{design_IC}
B_{a}(\rho)= L^{1-a}d^{a}\mathcal{D}_{d}^{(a)}F_{a}(\rho)\leq L^{1-a}d^{a}\mathcal{D}_{d}^{(a)}=:B_{a},
\end{align}
where $a=2,\cdots,t$ and the state-independent upper bound $B_{a}$ on IC is achieved at all pure states. As a typical example, $d+1$ MUBs of $\mathcal{H}_d$ form a quantum 2-design, and the corresponding total second-order IC is $\sum_{\theta=1}^{d+1}\sum_{i=1}^{d}p_{i|\theta}^2=1+tr(\rho^2)$ \cite{ref_43,multiple_1}.
Next, our discussions on EURs for DSMs will revolve around \myeq{design_IC}.

\section{ENTROPIC UNCERTAINTY RELATIONS}\label{sec_3}

The Rényi entropy of a probability distribution $P=(p_1,p_2,\cdots)$ is defined as
\begin{equation}
H_\alpha(P)=\frac{1}{1-\alpha}\ln\Big(\sum_{i}p_{i}^\alpha\Big),
\end{equation}
where $\alpha>0$ and $\alpha\neq1$. In the limit $\alpha\to1$, it converges to the Shannon entropy $H=-\sum_ip_i\ln p_i$.
Next, our focus will be on two particular forms of length-$L$ probability distributions defined below.

\begin{figure}[b]
\centering
\includegraphics[width=8.6cm]{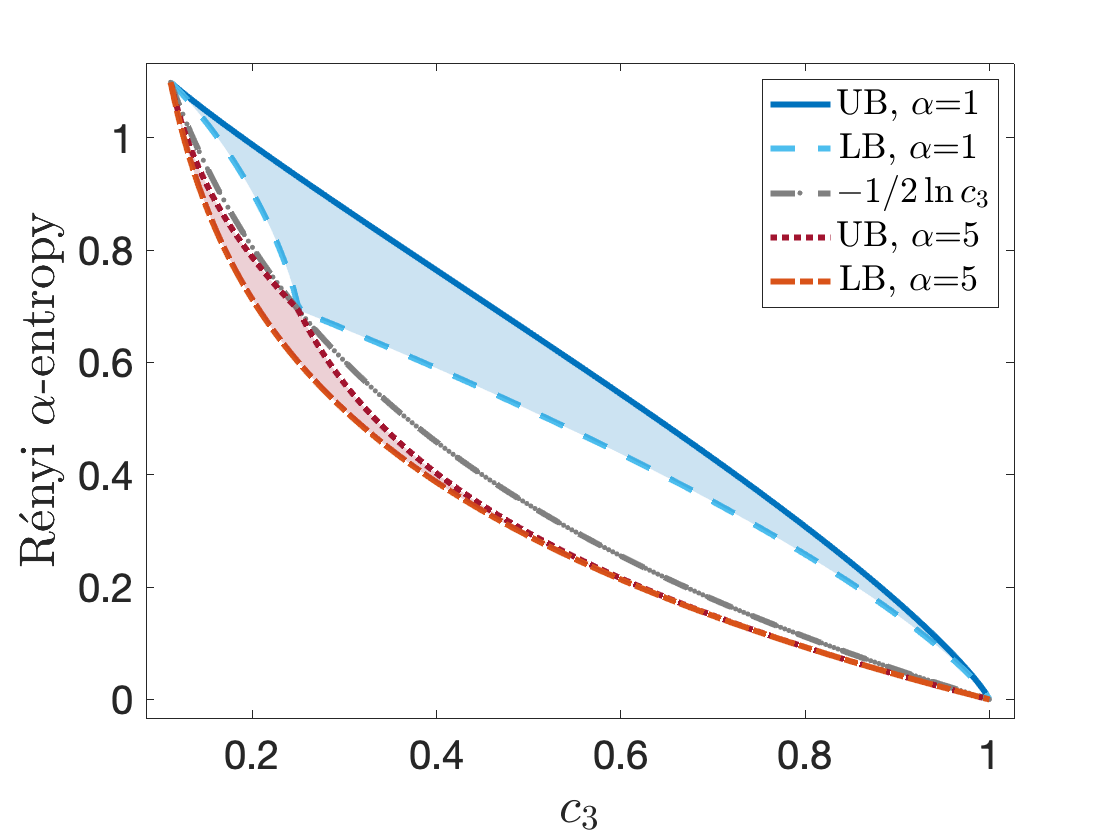}
\caption{The IC-entropy diagrams for the Shannon entropy ($\alpha=1$, blue region) and Rényi 5-entropy ($\alpha=5$, red region) of length-3 probability distributions.
According to Theorem 1 \eqref{thm_1},  the upper boundary (UB) and lower boundary (LB) of the IC-entropy diagram are respectively $H_\alpha(P_{x}^3[c_{3}])$ \eqref{px} and $H_\alpha(P_{y}[c_{3}])$ \eqref{py} when $\alpha=1$, whereas these two boundary functions are exchanged when $\alpha=5$.}
\label{information_diagram}
\end{figure}

\begin{align}
&P_{x}^{L}[c_{a}]=\big(\overbrace{p,\ p_{s},\ \cdots,\ p_{s}}^{L\ \text{probabilities}}\big), \label{px}\\
&\text{where}\ p^{a}+(L-1)p_{s}^{a}=c_{a},\ p\geq p_{s}=\frac{1-p}{L-1}\geq0;\nonumber\\
&P_{y}[c_{a}]=\big(\overbrace{p,\ \cdots,\ p,\ p_{s}}^{L'\ \text{probabilities}},\ 0,\ \cdots,\ 0\big), \label{py}\\
&\text{where}\ (L'-1)p^{a}+p_{s}^{a}=c_{a},\ p\geq p_{s}=1-(L'-1)p\geq0.\nonumber
\end{align}
In the above $a\geq2$ is an integer and $c_{a}\in[L^{1-a},1]$. 
It is noteworthy that  $I_{a}(P_{x}^{L}[c_{a}])=I_{a}(P_{y}[c_{a}])=c_{a}$ by definition. 
Moreover, $P_{x}^{L}[c_{a}]$ depends simultaneously on the value of $c_{a},\ a$, and $L$. Meanwhile, $P_{y}[c_{a}]$ is uniquely determined by $c_{a}$ and $a$, since this form of probability distributions must satisfy $(L')^{1-a}\leq c_{a}\leq(L'-1)^{1-a}$, therefore $L'=\big\lceil c_{a}^{1/(1-a)}\big\rceil$ is the round-up of $c_{a}^{1/(1-a)}$ to the nearest integer, which is not an independent variable.
We prove in \myapp{proof_of_theorem_1} the following.

\emph{Theorem 1.}
The Rényi $\alpha$-entropy $H_{\alpha}$ of an arbitrary length-$L$ probability distribution $P$ satisfies
\begin{align}
&H_\alpha(P_{x}^{L}[I_{a}(P)])\geq H_{\alpha}(P)\geq H_\alpha(P_{y}[I_{a}(P)])\ \text{if $\alpha\leq a$}
;\nonumber\\
&H_\alpha(P_{y}[I_{a}(P)])\geq H_{\alpha}(P)\geq H_\alpha(P_{x}^{L}[I_{a}(P)])\ \text{if $\alpha\geq a$}. \label{thm_1}
\end{align}
Theorem 1 describes the range of the R\'{e}nyi entropies in terms of the $a$th order IC of probability distributions.
We present in \myfig{information_diagram} the IC-entropy diagrams for the Shannon entropy $(\alpha=1)$ and R\'{e}nyi 5-entropy $(\alpha=5)$ of length-3 probability distributions when $a=3$.
Note in this case, $H_{3}(P)=-1/2\ln \big(I_3(P)\big)=-1/2 \ln c_{3}$.

We are also interested in lower bounds on the average entropy when considering multiple probability distributions. However, the boundary functions above cannot be utilized to address this case directly.  We instead introduce the function 
\begin{align}\label{Q_alpha}
\begin{split}
Q_{\alpha}(L,c_{a})=&\frac{\ln p^{\alpha}}{1-\alpha}+\frac{\ln L}{(1-\alpha)\ln [1+(L-1)^\frac{a}{\alpha}]}\\
&\times\ln\Big[1+(L-1)^\frac{a}{\alpha}\cdot\big(p_{s}/p\big)^{a}\Big]
\end{split}
\end{align}
to estimate the lower bound function $H_{\alpha}(P_{x}^{L}[c_{a}])$ when $\alpha\geq a$. Like \myeq{px}, here $p\in[\frac{1}{L},1]$ is the solution to the equation $p^{a}+(L-1)(\frac{1-p}{L-1})^{a}=c_{a}$, and $p_{s}=\frac{1-p}{L-1}$. Importantly,  $Q_{\alpha}(L,c_{a})\leq H_{\alpha}(P_{x}^{L}[c_{a}])$ and \myeq{Q_alpha} is convex with respect to $c_{a}$ (see \myapp{proof_of_theorem_2} for details), which immediately leads us to the theorem below.

\emph{Theorem 2.} 
When $\alpha\geq2$, the average Rényi $\alpha$-entropy over an arbitrary set of length-$L$ probability distributions $\{P_{\theta}\}_{\theta=1}^\Theta$ satisfies
\begin{equation}
\frac{1}{\Theta}\sum_{\theta}H_\alpha(P_{\theta})\geq Q_{\alpha}\big(L,c_{a}\big)=:q_2
\label{thm_2}
\end{equation}
for any integer $a\in[2,\alpha]$, where $c_{a}=\frac{1}{\Theta}\sum_{\theta}I_{a}(P_{\theta})$ is the average $a$th order IC.

Combining the IC \eqref{design_IC} of probability distributions induced by performing DSMs on quantum systems with Theorem 1 \eqref{thm_1} or Theorem 2 \eqref{thm_2}, it is already obvious to obtain EURs for  general DSMs. In an earlier work \cite{ket}, Ketterer and Gühne showed that for a number $\Theta$ of $L$-outcome measurements constructed from a quantum $t$-design, the respective average Rényi $\alpha$-entropy is lower bounded by 
\begin{equation}
q_{\rm Ket}=\frac{\alpha}{a(1-\alpha)}
\ln\big(B_{a}(\rho)\big)
\label{thm_ket}
\end{equation}
when $\alpha\geq a$ for any integer $a$ in the range $[2,t]$. Here, $B_{a}(\rho)$ is the average $a$th order IC in \myeq{design_IC}.
Following \mycite{ket}, Rastegin proposed another entropic bound \cite{ras} 
\begin{equation}\label{thm_ras}
q_{\rm Ras}=\frac{1}{1-\alpha}\Big[(\alpha-a)\ln p+\ln\big(B_{a}(\rho)\big)\Big],
\end{equation}
where $p\in[\frac{1}{L},1]$ is the solution to the equations $p^{a}+(L-1)(\frac{1-p}{L-1})^{a}=B_{a}(\rho)$. 
Rastegin also showed through examples that  $q_{\rm Ras}$ can be stronger than $q_{\rm Ket}$. 
We prove this in Appendix. \ref{com_bound}  that when $\alpha\geq a\geq2$, there is always  $q_{2}\geq q_{\rm Ras}\geq q_{\rm Ket}$. 
Additionally, it can be easily checked that $q_{2}=q_{\rm Ras}=q_{\rm Ket}=\frac{1}{1-a}\ln\big(B_{a}(\rho)\big)$ holds when $\alpha=a$, whereas in the limit $\alpha\to \infty$, $q_{2}=q_{\rm Ras}=-\ln p\geq q_{\rm Ket}=-\frac{1}{a}\ln\big(B_{a}(\rho)\big)$.

\begin{figure}[t]
\centering
\includegraphics[width=8.6cm]{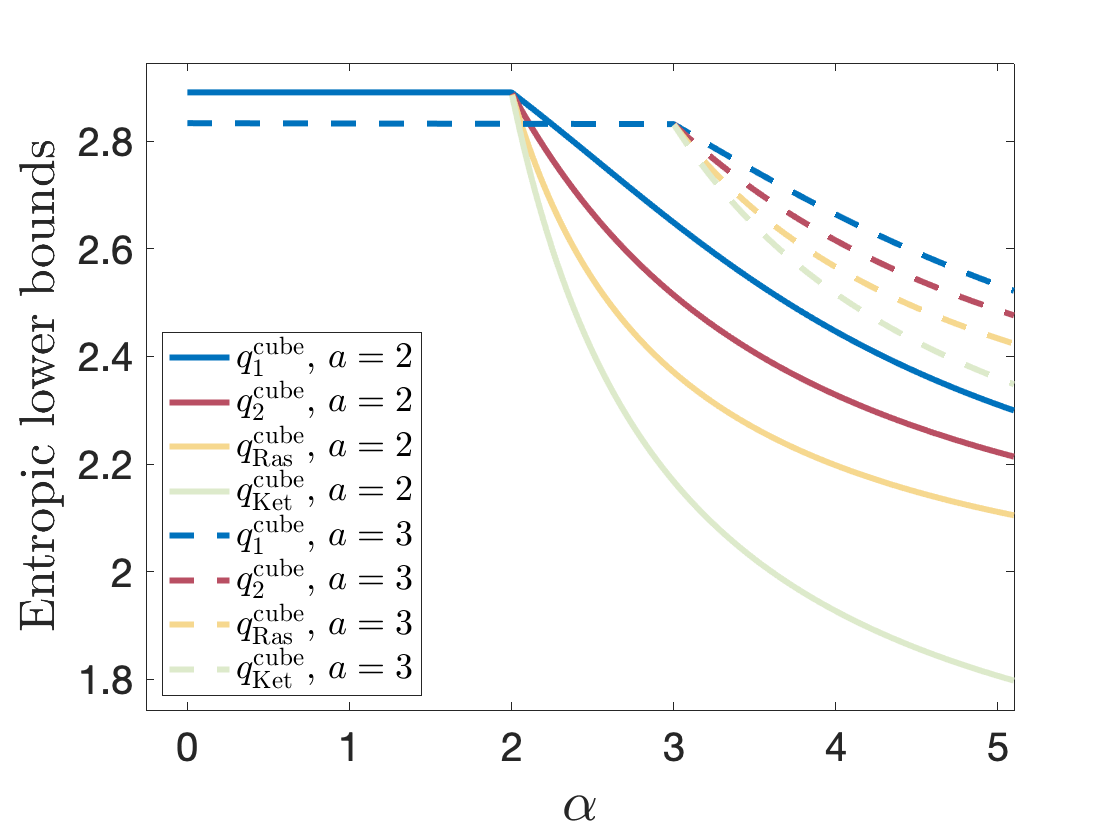}
\caption{State-independent entropic lower bounds for the 24-outcome POVM constructed from the deformed snub cube $7$-design: numerical comparisons between  $q_{1}^{\rm cube}$ \eqref{eur_cube}, $q_{2}^{\rm cube}$ \eqref{thm_2}, $q_{\rm Ras}^{\rm cube}$ \eqref{thm_ras} and $q_{\rm Ket}^{\rm cube}$ \eqref{thm_ket}.}\label{bound_cube}
\end{figure}

We remark here that the EURs given by Theorems 1 and 2 are strictly stronger than those proposed in \mycites{ket, ras} in general. As the first example, let us consider the quantum $7$-design consisting of $K=24$ unit vectors of $\mathcal{H}_2$ (this corresponds to 24 vertices, which form a deformed snub cube on the Bloch sphere \cite{ket}).  For the single POVM with $L=$24 possible measurement outcomes constructed from this quantum design, the corresponding state-independent upper bound on IC is then $B_{a}^{\rm cube}=\frac{24}{(a+1)12^{a}}$ \big(see \myeq{design_IC}\big). Combining it with Theorem 1 \eqref{thm_1}, we arrive at the state-independent entropic lower bounds below.
\begin{align}
q_{1}^{\rm cube}=\left\{
\begin{aligned}
&H_{\alpha}(P_{y}[B_{a}^{\rm cube}])\hspace{2.3em} \text{($\alpha\leq a)$}; \\
&H_{\alpha}(P_{x}^{24}[B_{a}^{\rm cube}])\hspace{1.8em} \text{($\alpha\geq a$)},
\end{aligned}
\right.
\label{eur_cube}
\end{align}
with $a=2,\cdots,7$, we present in \myapp{specific} its additional calculations.
In \myfig{bound_cube}, we present numerical comparisons between the aforementioned entropic lower bounds for $a=2,3$.
As shown, when $\alpha\geq a$, our Theorems lead to improved entropic lower bounds $q_{1}^{\rm cube}$ \eqref{eur_cube} and  $q_{2}^{\rm cube}=Q_{\alpha}(24,B_{a}^{\rm cube})$ \eqref{thm_2} compared to  the ones $q_{\rm Ras}^{\rm cube}=\frac{1}{1-\alpha}\big[(\alpha-a)\ln p+\ln(B_{a}^{\rm cube})\big]$ \eqref{thm_ras} and  $q_{\rm Ket}^{\rm cube}= \frac{\alpha}{a(1-\alpha)}\ln(B_{a}^{\rm cube})$ \eqref{thm_ket} proposed in \mycite{ras} and \mycite{ket} respectively.  
Notably, our Theorem 1 is also applicable for $\alpha<a$. Roughly speaking,  $q_{1}^{\rm cube}$ is invariant as $\alpha$ varies within the range $[0,a]$.

\begin{figure}[t]
\centering
\includegraphics[width=8.6cm]{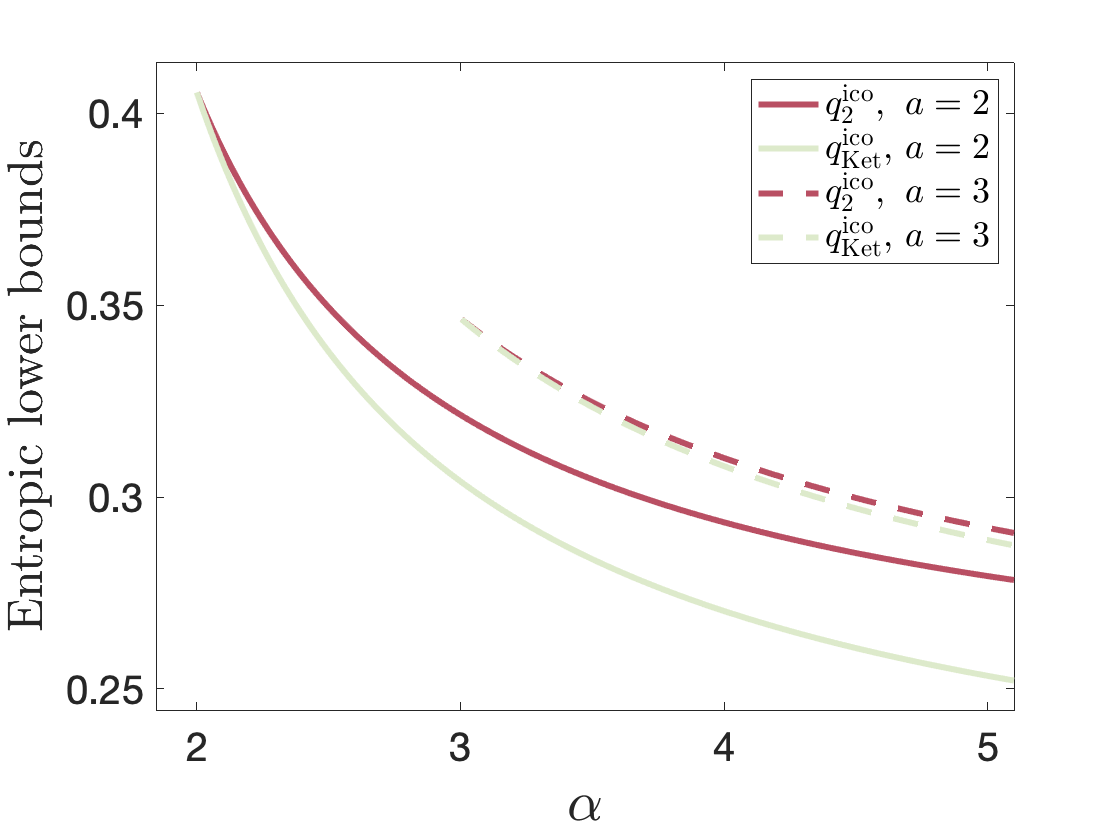}
\caption{State-independent entropic lower bounds for the six 2-outcome POVMs constructed from the icosahedron $5$-design: numerical comparisons between $q_{2}^{\rm ico}$ \eqref{eur_ico} and $q_{\rm Ket}^{\rm ico}$ \eqref{thm_ket}.}\label{bound_ico}
\end{figure}

We proceed to consider the quantum $5$-design consisting of $K=12$ unit vectors of $\mathcal{H}_{2}$ (this corresponds to 12 vertices which form an icosahedron on the Bloch sphere \cite{ket}). 
For this quantum design, we can construct $\Theta=6$ POVMs, each having $L=$2 possible measurement outcomes, and their average upper bound on IC is $B_{a}^{\rm ico}=\frac{2}{a+1}$ \big(see \myeq{design_IC}\big).
Applying Theorem 2 \eqref{thm_2}, we arrive at the state-independent entropic lower bounds
\begin{equation}
q_{2}^{\rm ico}=
\frac{1}{1-\alpha}\ln\big[p^{\alpha}+p^{\alpha-a}(1-p)^{a}\big]\ (\alpha\geq a),
\label{eur_ico}
\end{equation}
where $a=2,\cdots,5$ and $p\in[0.5,1]$ is the solution to the equation $p^{a}+(1-p)^{a}=\frac{2}{a+1}$.
We checked numerically that the maximum value of $q_{2}^{\rm ico}$ is attained at $a=2$ for $2\leq\alpha\leq3$ and at $a=3$ for $\alpha>3$, namely, 
\begin{align}
&\max_{a}\{q_{2}^{\rm ico}\}\nonumber \\
=&\left\{
\begin{aligned}
&\frac{1}{1-\alpha}\ln\big[p^{\alpha}+p^{\alpha-2}(1-p)^{2}\big]\ \hspace{1em}\text{($2\leq\alpha<3$)}; \\
&\frac{1}{1-\alpha}\ln\big[p^{\alpha}+p^{\alpha-3}(1-p)^{3}\big]\ \hspace{1em}\text{($\alpha\geq3$)}.
\end{aligned}
\right.\label{eur_ico_tight}
\end{align}
In the above $p=\frac{\sqrt{3}+1}{2\sqrt{3}}$. 
Our entropic lower bound $q_{2}^{\rm ico}$ \eqref{eur_ico} is compared in \myfig{bound_ico} with an earlier result  $q_{\rm Ket}^{\rm ico}=\frac{\alpha}{a(1-\alpha)}\ln(\frac{2}{a+1})$  \eqref{thm_ket} obtained by Ketterer \emph{et al.}  \cite{ket} for $a=2,3$ respectively.
As shown, $q_{2}^{\rm ico}$ is improved than $q_{\rm Ket}^{\rm ico}$, especially for large $\alpha$. We remark that $q_{2}$ \eqref{thm_2} is  equivalent to $q_{\rm Ras}$ \eqref{thm_ras} in this simple example of two-outcome DSMs, but  $q_{2}$ would generally be stronger than $q_{\rm Ras}$ for multiple-outcome DSMs, i.e., when $L>2$.

\section{ENTANGLEMENT DETECTION}\label{sec_4}
Quantum entanglement is an expensive resource involved in many quantum tasks, including quantum cryptographic protocols \cite{entangle_cryp_1, entangle_cryp_2, entangle_cryp_3}, and certified quantum random number generation \cite{entangle_rand}. 
These tasks frequently necessitate a highly efficient and convenient approach to verify whether an unknown quantum state is entangled and whether an entangled state remains entangled in the presence of environmental noise. 
Any separable quantum state can be written in the form $\rho_{\rm sep}=\sum_{k}p_{k}\rho_{k}^{(1)}\otimes\cdots\otimes\rho^{(N)}_k$, which describes an $N$-partite system that is, with probability $p_k$, in the product state $\rho_{k}^{(1)}\otimes\cdots\otimes\rho^{(N)}_k$, where $\{\rho_{k}^{(n)}\}_k$ are the local states describing the $n$th ($1\leq n\leq N$) subsystem. Correspondingly, non-separable states are said to be entangled.

Inspired by the entanglement criteria \cite{ED_source_and_quantity, ED_use,huang_new}  for bipartite systems, now we show how to detect multi-partite entanglement based on DSMs. 
For simplicity, we restrict ourselves to measurements that are already design-structured as individual POVMs. 
Our approach is then to perform, depending on the value of a local random variable drawn from some probability distribution $w_{\theta}^{(n)}$, the $\theta$th local DSM $\mathcal{M}_{\theta}^{(n)}$ on the $n$th subsystem. 
To be specific, we  introduce the correlation measure $J(\rho)=\sum_{i,\theta}tr(J_{i|\theta}\rho)$ for $N$-partite quantum states $\rho$, where $J_{i|\theta}=w_{\theta}^{(1)}{M}_{i|\theta}^{(1)}\otimes\cdots\otimes w_{\theta}^{(N)}{M}_{i|\theta}^{(N)}$ $(w_{\theta}^{(n)}>0, \sum_{\theta}w_{\theta}^{(n)}=1)$ is the correlation detection operator. 
Here, $M^{(n)}_{i|\theta}$ denotes the $i$th POVM effect of the $\theta$th local DSM to be performed on the $n$th subsystem with selection probability $w_{\theta}^{(n)}$.  In this stage, we assume that the sums of possible outcomes over all local measurements on individual subsystems are identical.
We prove in \myapp{proof_of_thm_3_4} two necessary conditions for multi-partite quantum states to be separable.

\emph{Theorem 3.} When performing a local DSM constructed from $t_{\theta}^{(n)}$-design with probability $w_{\theta}^{(n)}$ on the $n$th subsystem of an arbitrary $N$-partite separable system $\rho_{\rm sep}$, for any positive integers $\{a_n\}$ satisfying $2\leq a_n\leq t_{\theta}^{(n)}$ and $\sum_{n=1}^N \frac{1}{a_n}=1$ it  holds that
\begin{align}\label{thm_3}
J(\rho_{\rm sep})\leq
\prod_{n}\Big[\sum_\theta (w_{\theta}^{(n)})^{a_{n}}{B}_{a_{n},\theta}^{(n)}\Big]^{\frac{1}{a_{n}}},
\end{align}
where $B_{a_{n},\theta}^{(n)}$ is the upper bound on the  $a_n$th order IC \eqref{design_IC} of the probability distribution induced by performing the $\theta$th DSM on the $n$th subsystem.

In the above, $B_{a_{n},\theta}^{(n)}$ is a certainty measure whose value depends on the specific DSM that we choose to perform. Theorem 3 thus says that the correlation between subsystems of a system in any separable state is not strong enough to break the certainty bound as given by the right-hand side of \myeq{thm_3}. Violation of \myeq{thm_3} immediately indicates the presence of entanglement.  
Interestingly, simply by restricting $a_{n}$ to be even, we can obtain a stronger criterion. 
To this end, we modify the correlation measure as $\widetilde{J}(\rho)=\sum_{i,\theta}\big|\mathrm{tr}(\widetilde{J}_{i|\theta}\rho)\big|$, where the correlation detection operator becomes $\widetilde{J}_{i|\theta}=w_{\theta}^{(1)}\widetilde{M}_{i|\theta}^{(1)}\otimes\cdots\otimes w_{\theta}^{(N)}\widetilde{M}_{i|\theta}^{(N)}$, with
$\widetilde{M}_{i|\theta}^{(n)}=M_{i|\theta}^{(n)}-\frac{1}{d_n}tr(M_{i|\theta}^{(n)})\mathbbm{1}_{d_n}^{(n)}=M_{i|\theta}^{(n)}-\frac{1}{K_{\theta}^{(n)}}\mathbbm{1}_{d_n}^{(n)}$ being traceless and $d_n$ the dimension of the $n$th subsystem.

\emph{Theorem 4.} When performing a local DSM constructed from $t_{\theta}^{(n)}$-design with probability $w_{\theta}^{(n)}$ on the $n$th subsystem of an arbitrary $N$-partite separable system $\rho_{\rm sep}$, for any positive even integers $\{a_n\}$ satisfying $2\leq a_n\leq t_{\theta}^{(n)}$ and $\sum_{n=1}^N \frac{1}{a_n}=1$ it  holds that
\begin{align}\label{thm_4}
\widetilde{J}(\rho_{\rm sep})\leq
\prod_{n}\Big[\sum_\theta (w_{\theta}^{(n)})^{a_{n}}\widetilde{B}_{a_{n},\theta}^{(n)}\Big]^{\frac{1}{a_{n}}},
\end{align}
where $\widetilde{B}_{a_{n},\theta}^{(n)}=\sum_{r=0}^{a_n}C_{a_n}^r\frac{1}{(-K_{\theta}^{(n)})^{a_n-r}}B_{r,\theta}^{(n)}$ is the modified upper bound on the $a_n$th order IC \eqref{design_IC} of the probability distribution induced by performing the $\theta$th DSM on the $n$th subsystem.

\begin{figure}[bt]
\centering
\includegraphics[width=8.6cm]{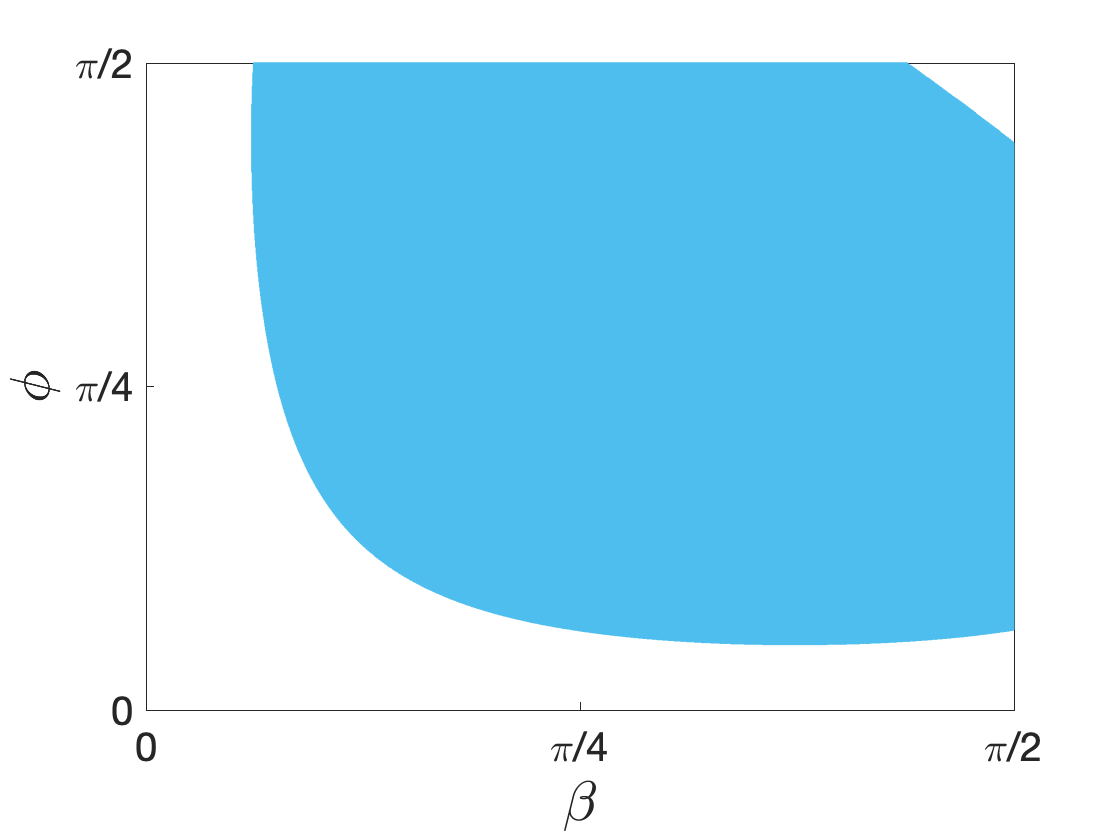}
\caption{Performance of Theorem 4 when performing the DSM constructed from the icosahedron 5-design on the state $\ket{\psi_{\beta,\phi}}$ \eqref{psi_beta_phi}, where the blue region necessarily indicates the violation of \myeq{thm_4}. }\label{ED_W_like}
\end{figure}

Now, we examine the effectiveness of Theorems 4 \eqref{thm_4} by adopting the DSM constructed from the icosahedron 5-design. We focus on detecting entanglement in $4$-qubit systems, with $a_{n}=4$ and $w_{\theta}^{(n)}=1$. 
The first example is some pure entangled states, given by
\begin{equation}\label{psi_beta_phi}
\begin{split}
\ket{\psi_{\beta,\phi}}=&\sin(\beta)\sin(\phi)\ket{0000}+\cos(\beta)\ket{1100}
\\+&\sin(\beta)\cos(\phi)\ket{1010},    
\end{split}    
\end{equation}
with $\beta,\phi\in(0,\frac{\pi}{2})$.  
According to \myeq{thm_4}, most of $\ket{\psi_{\beta,\phi}}$ are detected, but the detection fails when either  $\beta$ or $\phi$ are close to $0$, or when both $\beta$ and $\phi$ are close to $\frac{\pi}{2}$, as illustrated in \myfig{ED_W_like}.

\begin{figure}[bt]
\centering
\includegraphics[width=8.6cm]{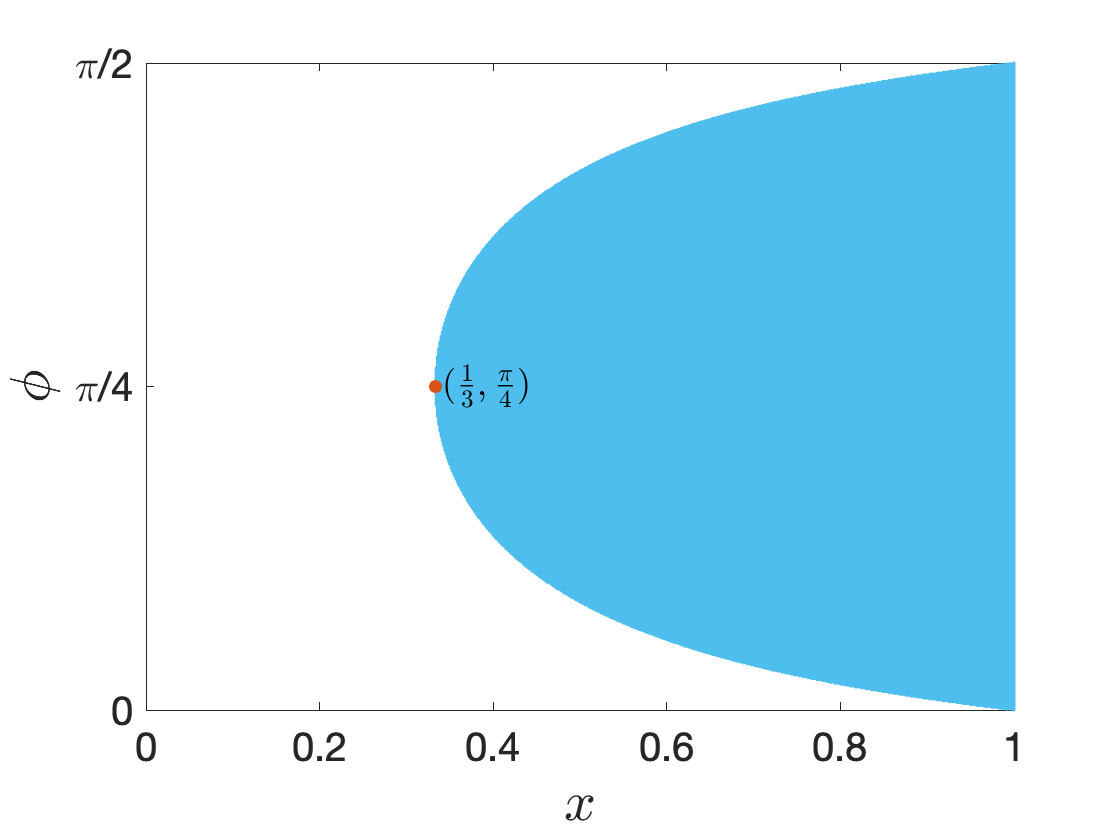}
\caption{Performance of Theorem 4 when performing the DSM constructed from the icosahedron 5-design on the state $\rho_{x,\phi}$ \eqref{rho_x_phi}, where the blue region necessarily indicates the violation of \myeq{thm_4}. }\label{ED_mixture}
\end{figure}

As a second example, we move on to consider the entanglement of an entangled state in the presence of white noise,
\begin{equation}\label{rho_x_phi}
    \rho_{x,\phi}=x\ket{\psi_{\phi}}\bra{\psi_{\phi}}+(1-x)\frac{\mathbbm{1}_{16}}{16},
\end{equation}
where $\ket{\psi_{\phi}}=\sin{\phi}\ket{0000}+\cos{\phi}\ket{1111}$ ($\phi\in[0,\frac{\pi}{2}]$) is a pure entangled state of four qubits (note $\ket{\psi_{\phi}}$ becomes maximally entangled at $\phi=\frac{\pi}{4}$) and $x\in[0,1]$ describes the amount of white noise.
In the case $x=1$, $\rho_{x,\phi}$  is entangled whenever $0<\phi<\frac{\pi}{2}$, but $\rho_{x,\phi}$ may cease to be entangled because of stronger noise, i.e., smaller $x$.  We utilize \myeq{thm_4} to test the entanglement of $\rho_{x,\phi}$. 
As  depicted in \myfig{ED_mixture},  \myeq{thm_4}  successfully detects all the entanglement of four-qubit states in the form of  $\rho_{x,\phi}$ \eqref{rho_x_phi} when $x=1$. 
But as the value of $x$ becomes smaller and the white noise increases, the detectable entanglement given by \myeq{thm_4} decreases.
And when $x<\frac{1}{3}$, no entanglement can be detected by \myeq{thm_4}.

\section{DISCUSSIONS AND CONCLUSION}\label{sec_5}
In the aforementioned examples of $\rho_{x,\phi}$ \eqref{rho_x_phi} and $\ket{\psi_{\beta,\phi}}$ \eqref{psi_beta_phi}, each subsystem adopts only one specific measurement (i.e., $w_{\theta}^{(n)}=1)$, and the measurements and integers $a_{n}$ are identical across various subsystems. 
We anticipate that by selecting appropriate measurements and integers $a_{n}$ for these subsystems, and adjusting their selection probabilities $w_{\theta}^{(n)}$, the performance of entanglement detection in Eqs.\hspace{0.3em}(\ref{thm_3},\ref{thm_4}) will be improved.
Also, we are interested in how entanglement detection can be impacted by the unbiasedness of measurements. Remarkably, it is found that the greater degree of unbiasedness corresponds to better detection performance, as detailed in \myapp{unbiasedness_}.
Furthermore, it is worth noting that all our theorems are based on the $a$th order IC of DSMs in \myeq{design_IC}, where $a$ is an integer within the range $[2,t]$.
We aspire to extend this range in the hope of obtaining more favorable results based on our theorems, a discussion of which can be found in \myapp{high_IC}.

In conclusion, we begin by exploring the particular forms of a single probability distribution that, conditioned on the value of IC, minimize or maximize the corresponding Rényi entropy. 
These particular forms allow us to establish an entropic lower bound for a single probability distribution. 
Subsequently, we formulate a function that is smaller than this lower bound and convex, serving as an entropic lower bound for multiple probability distributions. 
By combining these results with the IC of DSMs, we derive improved EURs for DSMs (this approach should also apply to the Tsallis entropy \cite{tsallis}).
Improvements to EURs would elevate their effectiveness in detecting quantum entanglement \cite{entanglement, huang_old}, and entropic steering test \cite{steering, ket, huang_new_}.
Furthermore,  we introduce the necessary criteria for multi-partite separable states, specifically, the inequalities between the correlation measure and the upper bounds on IC associated with DSMs. These criteria are demonstrated to be effective for entanglement detection through specific examples.

\acknowledgments
This work is supported by the National Natural Science Foundation of China (Grant No. 12175104) and the Innovation Program for Quantum Science and Technology (2021ZD0301701).

\appendix

\begin{widetext}

\section{PROOF OF THEOREM 1}\label{proof_of_theorem_1}
Given a length-$L$ probability distribution $P=(p_1,p_2,\cdots)$ under two constraints $\sum_{i}p_i=1$ and $I_{a}(P)=\sum_{i}p_i^a=c_{a}$, consider its Rényi entropy $H_\alpha(P)=\frac{1}{1-\alpha}\ln\big(\sum_{i}p_{i}^\alpha\big)$ at the extreme.

For simplicity, let us first consider the extreme of $\sum_{i}p_i^\alpha$. Using the Lagrange multiplier method, with $\lambda_0$ and $\lambda$ being multipliers, we obtain
\begin{equation}
\mathcal{L}=\sum_{i}p_i^\alpha+\lambda_0(\sum_{i}p_i-1)+\lambda(\sum_{i}p_i^a-c_{a}),\hspace{1em}
\frac{\partial{\mathcal{L}}}{\partial{p_{i}}}=\alpha p_i^{\alpha-1}+\lambda_0+\lambda a p_i^{a-1}=0.
\label{eq:equation1}
\end{equation}
Let $x_i=p_i^{a-1}$ and $f(x_i)=x_i^{\frac {\alpha -1}{a-1}}=p_i^{\alpha-1}$, calculate the second derivative of $f(x_i)$,
\begin{equation}
\frac{\mathrm{d}^{2}f(x_{i})}{\mathrm{d} x_i^{2}}
=\frac{(\alpha-1)(\alpha-a)}{(a-1)^2}x_i^{\frac{\alpha -2a+1}{a-1}}.
\label{eq:equation2}
\end{equation}
It is evident that $f(x_{i})$ is either convex or concave when $a\neq1,\alpha\neq1,\alpha\neq a$.
So the equation $\alpha f(x_{i})=-\lambda_{0}-\lambda tx_{i}$ from \eqref{eq:equation1} has at most two distinct nonzero solutions, given the convexity or concavity of $\alpha f(x_{i})$ and the linear form of $-\lambda_{0}-\lambda ax_{i}$.

Therefore, there are at most two different values of nonzero probabilities.
We denote the bigger probability as $p$, the smaller as $p_s$, the number of probabilities being $p$ as $N$, and the total number of nonzero probabilities as $N_t$. The two constraints and Rényi entropy can be rewritten as
\begin{equation}
{N}{p}+(N_{t}-{N}){p_{s}}=1, \hspace{1em}{N}{p^a}+(N_{t}-{N}){p_{s}^a}=c_{a}, \hspace{1em}
H_\alpha=\frac{1}{1-\alpha}\ln[{N}p^\alpha+({N_{t}}-{N})p_{s}^\alpha],
\label{eq: constraints and entropy}
\end{equation}
differentiate and reorganize these equations, denoting $\frac{p_{s}}{p}=\eta\in[0,1]$, we obtain
\begin{equation}
\mathrm{d}{H_\alpha}=\frac{({{p}{p_{s}}})^{\frac {\alpha+a-1}{2}}}{[{N}p^\alpha+({N_{t}}-{N})p_{s}^\alpha]a(p^{a-1}-p_{s}^{a-1})}
({g_{N_{t}}}\mathrm{d}N_{t}+g_{N}\mathrm{d}N),
\label{eq:differential entropy}
\end{equation}
\begin{equation}
{g_{N_{t}}}=\frac{1}{\alpha-1}
\big[(a-\alpha)\eta^\frac{\alpha+a-1}{2}
+a(\alpha-1)\eta^\frac{\alpha-a+1}{2}
-\alpha(a-1)\eta^\frac{a-\alpha+1}{2}\big],
\label{eq:fN}
\end{equation}
\begin{equation}
{g_{N}}=\frac{-1}{\alpha-1}
\big[(a-\alpha)(\eta^\frac{\alpha+a-1}{2}+\eta^{-\frac{\alpha+a-1}{2}})
+a(\alpha-1)(\eta^\frac{\alpha-a+1}{2}+\eta^{-\frac{\alpha-a+1}{2}})
-\alpha(a-1)(\eta^\frac{a-\alpha+1}{2}+\eta^{-\frac{a-\alpha+1}{2}})\big].
\label{eq:fNa}
\end{equation}
If $\alpha<a$, ${g_{N_{t}}}\geq0$, ${g_{N}}\leq0$, which leads to $\frac{\partial H_\alpha}{\partial N_{t}}\geq0$, $\frac{\partial H_\alpha}{\partial {N}}\leq0$; if $\alpha>a$, ${g_{N_{t}}}\leq0$, ${g_{N}}\geq0$, which leads to $\frac{\partial H_\alpha}{\partial N_{t}}\leq0$, $\frac{\partial H_\alpha}{\partial N}\geq0$.
Therefore, $H_{\alpha}$ reaches its extreme if $N_{t}$ and $N$ are maximal or minimal.
Obviously, $N_{t}=L$ when it is maximal.
And according to $c_{a}\geq N_{t}^{1-a}$, $N_{t}=\big\lceil c_{a}^{1/(1-a)}\big\rceil$ when it is minimal, denote it as $L'$.  

So in conclusion,
\begin{equation}
\frac{\ln \big[p^{\alpha}+(L-1)p_{s}^{\alpha}\big]}{1-\alpha}
\geq H_{\alpha}(P) \geq
\frac{\ln\big[(L'-1)p^{\alpha}+p_{s}^{\alpha}\big]}{1-\alpha},
\end{equation}
when $\alpha \leq a$,
\begin{equation}
\frac{\ln\big[(L'-1)p^{\alpha}+p_{s}^{\alpha}\big]}{1-\alpha}
\geq H_{\alpha}(P)\geq
 \frac{\ln \big[p^{\alpha}+(L-1)p_{s}^{\alpha}\big]}{1-\alpha},
\end{equation}
when $\alpha \geq a$. 
The value of $p,p_{s}$ can be obtained through a numerical approach under the two constraints in \eqref{eq: constraints and entropy}.
Thus, we complete the proof of Theorem 1 \eqref{thm_1}.

\section{PROOF OF THEOREM 2}\label{proof_of_theorem_2}
We begin by proving the convexity of $Q_{\alpha}(L,c_{a})$ \eqref{Q_alpha} with respect to $c_{a}$. Rewrite $Q_{\alpha}(L,c_{a})$ as
\begin{align}
\bigg\{\alpha-\frac{a\ln L}{\ln \big[1+(L-1)^\frac{a}{\alpha}\big]}\bigg\}\frac{\ln {p}}{1-\alpha}
+\frac{\ln L}{\ln\big[1+(L-1)^\frac{a}{\alpha}\big]}\frac{1}{1-\alpha}
\ln\big[p^{a}+(L-1)^\frac{a}{\alpha}p_{s}^{a}\big],
\label{align:convexity}
\end{align}	
we note that $\alpha-\frac{a\ln L}{\ln [1+(L-1)^\frac{a}{\alpha}]}$, $\frac{\ln L}{\ln[1+(L-1)^\frac{a}{\alpha}]}$ are non-negative, and $\frac{\ln {p}}{1-\alpha}$ is convex. Taking the derivative of the remaining part $\frac{1}{1-\alpha}\ln\big[p^a+(L-1)^\frac{a}{\alpha}p_{s}^a\big]$ with respect to $c_{a}$, we obtain
\begin{equation}
-\frac{1}{\alpha-1}\frac{1}{p^a+(L-1)^\frac{a}{\alpha}p_{s}^a}
\bigg\lbrace1+\frac{\big[1-(L-1)^{\frac{a-\alpha}{\alpha}}\big]p_{s}^{a-1}}{p^{a-1}-p_{s}^{a-1}}\bigg\rbrace, 
\label{eq:convexity}
\end{equation}
where both $\frac{1}{p^a+(L-1)^\frac{a}{\alpha}p_{s}^a}$ and $1+\frac{[1-(L-1)^{\frac{a-\alpha}{\alpha}}]p_{s}^{a-1}}{p^{a-1}-p_{s}^{a-1}}$ are non-negative and monotonic decreasing with respect to $c_{a}$. Therefore, $\frac{1}{1-\alpha}\ln\big[p^a+(L-1)^\frac{a}{\alpha}p_{s}^a\big]$ is also convex.
Hence, the whole formula $Q_{\alpha}(L, c_{a})$ is convex, and its convexity implies that
\begin{equation}\label{thm_2_1}
\frac{1}{\Theta}\sum_{\theta}Q_{\alpha}(L,c_{a,\theta})
\geq Q_{\alpha}(L,\frac{1}{\Theta}\sum_{\theta}c_{a,\theta}).
\end{equation}

Then, we prove that
\begin{equation}\label{thm_2_2}
\frac{\ln\big[p^{\alpha}+(L-1)p_{s}^{\alpha}\big]}{1-\alpha}\geq Q_{\alpha}(L,c_{a}),
\end{equation}
the difference between them can be written as
\begin{align}
\frac{1}{\alpha-1}
\bigg\lbrace
{\ln L}\frac{\ln\big[1+(L-1)^\frac{a}{\alpha}(\frac{p_{s}}{p})^a\big]}{\ln\big[1+(L-1)^\frac{a}{\alpha}\big]}-\ln\big[1+(L-1)(\frac{p_{s}}{p})^\alpha\big]\bigg\rbrace, 
\end{align}
which equals to $0$ when $\alpha=a$, larger than 0 when $\alpha>a$ since $\frac{\ln[1+(L-1)^\frac{a}{\alpha}(\frac{p_{s}}{p})^a]}{\ln[1+(L-1)^\frac{a}{\alpha}]}$ is monotonic decreasing with respect to $a$.

And Theorem 1 \eqref{thm_1} guarantees that when $\alpha\geq a$,
\begin{equation}\label{thm_2_3}
H_\alpha(P)
\geq \frac{\ln\big[p^{\alpha}+(L-1)p_{s}^{\alpha}\big]}{1-\alpha}.
\end{equation}

Combining Eqs.\hspace{0.3em}(\ref{thm_2_1}, \ref{thm_2_2}, \ref{thm_2_3}), we complete the proof of Theorem 2 \eqref{thm_2}.

\section{COMPARISONS BETWEEN ENTROPIC LOWER BOUNDS}\label{com_bound}
This section shows that our entropic lower bound $q_2$  \eqref{thm_2} would  never be weaker than previous ones. As a complement to the numerical results presented in figures (\ref{bound_cube}, \ref{bound_ico}), here we compare $q_2$ with Ketterer \emph{et al.}'s bound $q_{\rm Ket}$ \eqref{thm_ket} and Rastegin's bound $q_{\rm Ras}$ \eqref{thm_ras} analytically.
\begin{equation}\left\{
\begin{aligned}
q_{\rm Ket}&=\frac{\alpha}{(1-\alpha)a}\ln c_{a},\\
q_{\rm Ras}&=\frac{1}{1-\alpha}\big[(\alpha-a)\ln p+\ln c_{a}\big],\\
q_{2}&=\frac{\alpha}{1-\alpha}\ln p+\frac{1}{1-\alpha}\frac{\ln L}{\ln\big[1+(L-1)^{a/\alpha}\big]}\ln\big[1+(L-1)^{a/\alpha}(p_s/p)^a\big].
\end{aligned}
\right.\end{equation}
To simplify the analysis, let us consider the following functions of $\alpha$ where $a,c_{a}, L$ are constants:
\begin{equation}\left\{
\begin{aligned}
f_1(\alpha)&=-\frac{\alpha}{a}\ln c_{a},\\
f_2(\alpha)&=-(\alpha-a)\ln p-\ln c_{a},\\
f_3(\alpha)&=-\alpha\ln p-\frac{\ln L}{\ln\big[1+(L-1)^{a/\alpha}\big]}\ln\big[1+(L-1)^{a/\alpha}(p_s/p)^a\big],\\
f(z)&=-\frac{\ln L}{\ln\big[1+(L-1)^{z}\big]}\ln\big[1+(L-1)^zx\big].
\end{aligned}
\right.\end{equation}
When $\alpha=a$ it is obvious that $q_{\rm Ket}=q_{\rm Ras}=q_{2}=\frac{1}{1-a}\ln c_{a}$, or equivalently $f_1(a)=f_2(a)=f_3(a)=-\ln c_{a}$. 
Therefore, we only need to consider the case when $\alpha>a$.
First, observe that $\frac{d f_2}{d \alpha}-\frac{d f_1}{d \alpha}=-\ln p+\frac{1}{a}\ln c_{a}\geq0$ since $c_{a}=p^a+(L-1)p_s^a\geq p^a$, which demonstrates $q_{\rm Ras}\geq q_{\rm Ket}$ when $\alpha\geq a\geq2$. 
Second, observe that  $f(z)$ is a monotonic decreasing function of $z\ (0<z\leq1)$ for any $x\in[0,1]$, which essentially means $f_3(\alpha)-f_2(\alpha)$ is an increasing function of $\alpha$, i.e., $q_{2}\geq q_{\rm Ras}$ when $\alpha\geq a\geq2$.
Therefore, we have $q_{2}\geq q_{\rm Ras}\geq q_{\rm Ket}$ when $\alpha\geq a\geq2$.

\section{ADDITIONAL CALCULATIONS}\label{specific}
For the single POVM with $L=$24 possible measurement outcomes constructed from the 7-design which forms a deformed snub cube on the Bloch sphere, we have
\begin{align*}
q_{1}^{\rm cube}=\left\{
\begin{aligned}
H_{\alpha}(P_{y}[B_{a}^{\rm cube}])\hspace{2em}  &\text{($\alpha\leq a)$}; \\
H_{\alpha}(P_{x}^{24}[B_{a}^{\rm cube}])\hspace{2em}  &\text{($\alpha\geq a$)},
\end{aligned}
\right.
\end{align*}
with $a=2,\cdots,7$ and $B_{a}^{\rm cube}=\frac{24}{(a+1)12^{a}}$. Particularly, it can be checked that $\lceil (B_a^{\rm cube})^{\frac{1}{1-a}}\rceil=18$ when $a=2$ and $\big\lceil (B_a^{\rm cube})^{\frac{1}{1-a}}\big\rceil=17$ when $a=3$. According to Eqs. (\ref{px}, \ref{py})  we have 

\begin{align}
q_{1}^{\rm cube}=\left\{
\begin{aligned}
&\frac{1}{1-\alpha}\ln\big[17p^{\alpha}+(1-17p)^{\alpha}\big]=\ln 18\approx 2.89\hspace{2.5em} \text{($\alpha\leq a=2)$}; \\
&\frac{1}{1-\alpha}\ln\big[16p^{\alpha}+(1-16p)^{\alpha}\big]\hspace{9em} \text{($\alpha\leq a=3)$}; \\
&\frac{1}{1-\alpha}\ln\Big[p^{\alpha}+23\Big(\frac{1-p}{23}\Big)^{\alpha}\Big]\hspace{9.3em} \text{($\alpha\geq a$)}.
\end{aligned}
\right.
\label{eur_cube_1}
\end{align}
 In the first equation,  $p=\frac{1}{18}$ is the solution to the equation $17p^{2}+(1-17p)^{2}=B_2^{\rm cube}=\frac{1}{18}$. This means, besides 6 zero-valued probabilities, $P_y[B_2^{\rm cube}]$ is a uniform distribution over 18 nonzero probabilities and $H_\alpha(P_y[B_2^{\rm cube}])\equiv\ln 18$ is independent of $\alpha$. 
 In the second equation,  $1/17<p=(53-\sqrt{21})/816<1/16$ is the solution to the equation $16p^{3}+(1-16p)^{3}=B_3^{\rm cube}=\frac{1}{288}$.  
 In the third equation, $p$ is the solution to the equation $p^{a}+23(\frac{1-p}{23})^{a}=B_a^{\rm cube}=\frac{24}{(a+1)12^{a}}$ in the range [$\frac{1}{24}$, 1] for any $a=2,\cdots,7$.

\section{PROOF OF THEOREMS 3 AND 4}\label{proof_of_thm_3_4}
On an $N$-partite system, a relationship exists between any separable state $\rho_{\rm sep}=\sum_{k}p_{k}\rho_{k}^{(1)}\otimes\cdots\otimes\rho_{k}^{(N)}$ ($p_k>0$, $\sum_{k}p_{k}=1$), product state $\rho_{\rm pro}=\rho_{k}^{(1)}\otimes\cdots\otimes\rho_{k}^{(N)}$ and their correlation detection operator $J_{i|\theta}=w_{\theta}^{(1)} M^{(1)}_{i|\theta}\otimes\cdots\otimes w_{\theta}^{(N)}M^{(N)}_{i|\theta}$ ($w_{\theta}^{(n)}>0, \sum_{\theta}w_{\theta}^{(n)}=1$):

\begin{equation}
\sum_{i,\theta}\mathrm{tr}\big(J_{i|\theta}\rho_{\rm sep}\big)
=\sum_{k}p_{k}\sum_{i,\theta}\mathrm{tr}\big(J_{i|\theta}\rho_{\rm pro}\big).
\end{equation}
For any product state, we have
\begin{align}
\sum_{i,\theta}\mathrm{tr}\big(J_{i|\theta}\rho_{\rm pro}\big)
=\sum_{i,\theta}\prod_{n}w_{\theta}^{(n)}p_{i|\theta}^{(n)}
\leq
\prod_{n}\Big[\sum_{i,\theta}(w_{\theta}^{(n)}p_{i|\theta}^{(n)})^{a_{n}}\Big]^{\frac{1}{a_{n}}},
\end{align} 
with $a_{n}>1$, $\sum_{n}\frac{1}{a_{n}}=1$, this is a direct consequence of Hölder's inequality,  and $p_{i|\theta}$ is the probability of obtaining the $i$th outcome when performing the $\theta$th measurement.
Also, note that
\begin{equation}
\prod_{n}\Big[\sum_{i,\theta}(w_{\theta}^{(n)} p_{i|\theta}^{(n)})^{a_{n}}\Big]^{\frac{1}{a_{n}}}
\leq\prod_{n}\Big[\sum_\theta (w_{\theta}^{(n)})^{a_{n}}{B}_{a_{n},\theta}^{(n)}\Big]^{\frac{1}{a_{n}}},
\end{equation}
where $B_{a_{n},\theta}^{(n)}$ is the upper bound on IC \eqref{design_IC} of the DSM on the $n$th  local subsystem.
Combining the above expressions, we complete the proof of Theorem 3 \eqref{thm_3}.

Theorem 4 is proved similarly.
The modified correlation detection operator is $\widetilde{J}_{i|\theta}=w_{\theta}^{(1)}\widetilde{M}_{i|\theta}^{(1)}\otimes\cdots\otimes w_{\theta}^{(N)}\widetilde{M}_{i|\theta}^{(N)}$ with $\widetilde{M}_{i|\theta}^{(n)}=M_{i|{\theta}}^{(n)}-\frac{1}{K_{\theta}^{(n)}}\mathbbm{1}_{d}^{(n)}$. For any separable and product state, there is
\begin{align}
\sum_{i,\theta}\Big|\mathrm{tr}
\big(\widetilde{J}_{i|\theta}\rho_{\rm sep}\big)\Big|
\leq\sum_{k}p_{k}\sum_{i,\theta}\Big|\mathrm{tr}\big(\widetilde{J}_{i|\theta}\rho_{\rm pro}\big)\Big|.
\end{align}
For any product state, we have
\begin{align}
\sum_{i,\theta}\Big|\mathrm{tr}\big(\widetilde{J}_{i|\theta}\rho_{\rm pro}\big)\Big|
=
\sum_{i,\theta}\prod_{n}\Big|w_{\theta}^{(n)}\widetilde{p}_{i|\theta}^{(n)}\Big|
\leq
\prod_{n}\Big(\sum_{i,\theta}\Big|w_{\theta}^{(n)}\widetilde{p}_{i|\theta}^{(n)}\Big|^{a_{n}}\Big)^{\frac{1}{a_{n}}}
\leq\prod_{n}\Big[\sum_{\theta}(w_{\theta}^{(n)})^{a_{n}}\widetilde{B}_{a_{n},\theta}^{(n)}\Big]^{\frac{1}{a_{n}}},
\end{align}
here, $a_{n}$ is restricted to an even number,  $\widetilde{p}_{i|\theta}^{(n)}=p_{i|\theta}^{(n)}-\frac{1}{K_{\theta}^{(n)}}$ and $\widetilde{B}_{a_{n},\theta}^{(n)}$ is the modified upper bound on IC.
Thus, we complete the proof of Theorem 4 \eqref{thm_4}.

\section{THE IMPACT OF UNBIASEDNESS ON ENTANGLEMENT DETECTION}\label{unbiasedness_}
The measure of unbiasedness for two bases $\{\ket{a_{i}}\}$ and $\{\ket{b_{i}}\}$ in $\mathcal{H}_{d}$ has been proposed in \mycite{unbiasedness} as: $U=d-1-\sum_{i,j}\big(|\braket{a_{i}|b_{j}}|^{2}-\frac{1}{d}\big)^{2}$. We extend this measure to multiple bases $\{\ket{a_{i}}\}, \{\ket{b_{i}}\},\cdots$ as:
\begin{equation}
U =(\Theta-1)(d-1)- \frac{2}{\Theta}
\sum_{a<b}\sum_{i,j} \Big(|\braket{a_{i}|b_{j}}|^{2}-\frac{1}{d}\Big)^{2}
,
\end{equation}
where $0\leq U\leq(\Theta-1)(d-1)$. This definition aligns well with our intuition, indicating that the eigenbases of compatible observables are entirely biased. In contrast, those of the complementary observables (i.e., MUBs) exhibit the greatest degree of unbiasedness. The complete set of MUBs reveals information about states completely.

Considering collections of three random bases with diverse unbiasedness in $\mathcal{H}_{2}$, we adopt them and their conjugate as measurements.
Utilizing their upper bound on IC from the results of \mycite{huang_new}, we apply Theorem 3 \eqref{thm_3} (which, in principle, applies to any measurements as long as their upper bound on IC is available) to detect entanglement in Isotropic states on $2$-qubit systems, with $a_{n}=2$ and $w_{\theta}^{(n)}=1$.
The Isotropic states are expressed as
\begin{equation}\label{rho_x}
\rho_{x}=x\ket{\Phi^{+}}\bra{\Phi^{+}}+(1-x)\frac{\mathbbm{1}_{4}}{4},  
\end{equation}
where $x\in[0,1]$ (with $\rho_{x}$ being entangled if $x>\frac{1}{3}$). Here, $\ket{\phi^{+}}=\frac{1}{\sqrt{2}}\big(\ket{00}+\ket{11}\big)$ is the maximally entangled state, and $\frac{\mathbbm{1}_{4}}{4}$ is the maximally mixed state. 
As depicted in \myfig{unbiasedness}, the performance of entanglement detection improves with greater unbiasedness, and applying MUBs successfully detects all entangled states.

\begin{figure}[bt]
\centering
\includegraphics[width=8.6cm]{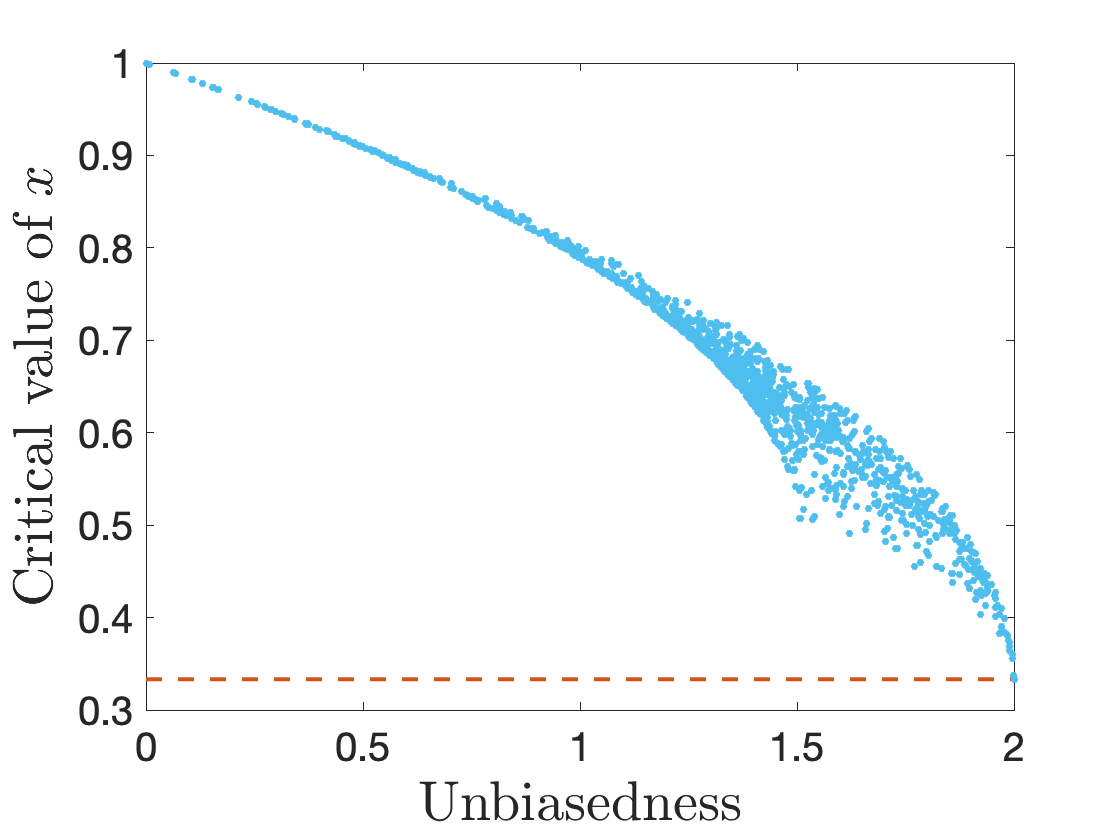}
\caption{Entanglement detection for $\rho_{x}$ using collections of three random bases with diverse unbiasedness in $\mathcal{H}_{2}$. The blue circle represents the critical value of $x$ in random bases: if $x$ exceeds it, $\rho_{x}$ \eqref{rho_x} is identified as entangled according to Theorem 3 \eqref{thm_3}. Therefore, the smaller the critical value, the better the performance of entanglement detection. $\rho_{x}$ is entangled if $x>\frac{1}{3}$ (red dashed line) intrinsically.}\label{unbiasedness}
\end{figure}

\section{HIGHER ORDER IC OF DSMS}\label{high_IC}
The Choi-Jamiołkowski isomorphism \cite{isomorphism} establishes a correspondence where any operator $A$ on $\mathcal{H}_{d}$ can be represented as a vector $\ket{A}$ in $\mathcal{H}_{d}\otimes\mathcal{H}_{d}$: $\ket{A}=A\otimes\mathbbm{1}_{d}\sum_{i}\ket{i}\otimes\ket{i}^{*}$.
Adopting this isomorphism and following \mycite{huang_new}, we introduce the generalized view operator associated with the DSMs $\{\mathcal{M}_{\theta}\}=\{M_{i|\theta}\}$: $G^{a}=\sum_{i,\theta}\ket{\widetilde{M}_{i|\theta}^{a}}\bra{\widetilde{M}_{i|\theta}^{a}}$, with $\widetilde{M}_{i|\theta}^{a}=M_{i|{\theta}}^{\otimes a}-\frac{1}{L^{a}}\mathbbm{1}_{d}^{\otimes a}$. Based on the generalized view operator, we propose an inequality providing an upper bound on the IC of DSMs:
\begin{align}\label{IC_high}
\sum_{\theta}I_{2a}(\mathcal{M}_{\theta})_{\rho}\leq
\big\|G^{a}\big\|\left[\big(\mathrm{tr}(\rho^{2})\big)^{a}+\frac{2F_{a}(\rho)-1}{d^{a}(1-h)} \right],
\end{align}
where $a$ is any integer in $[2,t]$, $F_{a}(\rho)=\mathrm{tr}(\mathbbm{1}_{d^{a}}^{\rm sym}\rho^{\otimes a})$, $h=\mathrm{tr}\big(\mathbbm{1}_{d^{a}}^{\rm sym}(\frac{\mathbbm{1}_{d}}{d})^{\otimes a}\big)=\frac{(a+d-1)!}{a!(d-1)!d^{a}}$, and $\|\cdot\|$ denotes the operator norm (i.e., the largest eigenvalue of the operator).

This inequality \eqref{IC_high} focuses on the IC of higher order, specifically even numbers in the range $[4,2t]$, for DSMs.
Given the increased options for IC, the EURs derived from Theorems 1 and 2 can potentially be improved.
This improvement is particularly relevant for the Rényi $\alpha-$entropy with large values of $\alpha$, including $\alpha\to\infty$ where better results can be obtained.
Furthermore, when applying Theorem 3 to detect entanglement, it requires measurements with available IC of higher order, a requirement that can be satisfied by the proposed inequality \eqref{IC_high}.
The following is its proof.

We define a ket $\ket{v}=\ket{\rho^{\otimes a}-\frac{1}{d^{a}}\mathbbm{1}_{d}^{\otimes a}}+\frac{1}{d^{a}(1-h)}\ket{\mathbbm{1}_{d^{a}}^{\rm sym}-h\mathbbm{1}_{d}^{\otimes a}}$. Given that in Choi-Jamiołkowski isomorphism for any operators $A_{1}$ and $A_{2}$: $\braket{A_{1}|A_{2}}=\mathrm{tr}(A_{1}^{\dagger}A_{2})$, and for DSMs there is $\sum_{i,\theta}\mathrm{tr}(\mathbbm{1}_{d^{a}}^{\rm sym}M_{i|\theta}^{\otimes a})=\sum_{i,\theta}\mathrm{tr}(M_{i|\theta}^{\otimes a})$, we proceed to calculate $\bra{v}G^{a}\ket{v}$ and $\braket{v|v}$:
\begin{align}
&\bra{v}G^{a}\ket{v}\nonumber\\
=&\sum_{i,\theta}\bigg\{\mathrm{tr}\Big[\big(\rho^{\otimes a}-\frac{1}{d^{a}}\mathbbm{1}_{d}^{\otimes a}\big)\big(M_{i|{\theta}}^{\otimes a}-\frac{1}{L^{a}}\mathbbm{1}_{d}^{\otimes a}\big)\Big]\bigg\}^{2}+\frac{1}{d^{2a}(1-h)^{2}}\bigg\{\mathrm{tr}\Big[\big(\mathbbm{1}_{d^{a}}^{\rm sym}-h\mathbbm{1}_{d}^{\otimes a}\big)\big(M_{i|{\theta}}^{\otimes a}-\frac{1}{L^{a}}\mathbbm{1}_{d}^{\otimes a}\big)\Big]\bigg\}^{2}\nonumber\\
&+\frac{2}{d^{a}(1-h)}\mathrm{tr}\Big[ \big(\rho^{\otimes a}-\frac{1}{d^{a}}\mathbbm{1}_{d}^{\otimes a}\big)\big(M_{i|{\theta}}^{\otimes a}-\frac{1}{L^{a}}\mathbbm{1}_{d}^{\otimes a}\big)\Big]\mathrm{tr}\Big[\big(\mathbbm{1}_{d^{a}}^{\rm sym}-h\mathbbm{1}_{d}^{\otimes a}\big)\big(M_{i|{\theta}}^{\otimes a}-\frac{1}{L^{a}}\mathbbm{1}_{d}^{\otimes a}\big) \Big]\nonumber\\
=&\sum_{i,\theta} \big(p_{i|\theta}^{a}-\frac{1}{L^{a}}\big)^{2}+\frac{1}{L^{2a}}+\frac{2}{L^{a}}\big(p_{i|\theta}^{a}-\frac{1}{L^{a}}\big)\nonumber\\
=&\sum_{i,\theta}p_{i|\theta}^{2a}.\label{D1}\\
&\braket{v|v}\nonumber\\
=&\mathrm{tr}\Big[(\rho^{\otimes a}-\frac{1}{d^{a}}\mathbbm{1}_{d}^{\otimes a})^{2}\Big]+\frac{1}{d^{2a}(1-h)^{2}}\mathrm{tr}\Big[(\mathbbm{1}_{d^{a}}^{\rm sym}-h\mathbbm{1}_{d}^{\otimes a})^{2}\Big]+\frac{2}{d^{a}(1-h)}\mathrm{tr}\Big[(\rho^{\otimes a}-\frac{1}{d^{a}}\mathbbm{1}_{d}^{\otimes a})(\mathbbm{1}_{d^{a}}^{\rm sym}-h\mathbbm{1}_{d}^{\otimes a})\Big]\nonumber\\
=&\big(\mathrm{tr}(\rho^{2})\big)^{a}+\frac{1}{d^{a}}-\frac{2}{d^{a}}+\frac{1}{d^{2a}(1-h)^{2}}\big(d^{a}h+d^{a}h^{2}-2d^{a}h^{2}\big)+\frac{2}{d^{a}(1-h)}\big(F_{a}(\rho)-h\big)\nonumber\\
=&\big(\mathrm{tr}(\rho^{2})\big)^{a}+\frac{2F_{a}(\rho)-1}{d^{a}(1-h)},\label{D2}
\end{align}
here $F_{a}(\rho)=\mathrm{tr}(\mathbbm{1}_{d^{a}}^{\rm sym}\rho^{\otimes a})$. Take the results in  \eqref{D1} and \eqref{D2} into $\bra{v}G^{a}\ket{v}\leq\big\|G^{a}\big\|\braket{v|v}$, we complete the proof of \myeq{IC_high}.

\end{widetext}


\begin{thebibliography}{99}

\bibitem{resource_1}
M. Oszmaniec and T. Biswas, 
\newblock Quantum \textbf{ 3}, 133 (2019).
%Operational Relevance of Resource Theories of Quantum Measurements

\bibitem{resource_2}
T. Guff, N. A. McMahon, Y. R. Sanders, and A. Gilchrist,
\newblock J. Phys. A: Math. Theor. \textbf{54}, 225301 (2021).
%A Resource Theory of Quantum Measurements

\bibitem{MUB_1}
I. Ivonovic,
\newblock J. Phys. A: Math. Gen. \textbf{14}, 3241 (1981).
%, Geometrical description of quantal state determination

\bibitem{MUB_2}
W. K. Wootters and B. D. Fields,
\newblock Ann. Phys. \textbf{191}, 363 (1989).
% \textit{Optimal State-Determination by Mutually Unbiased Measurements},


\bibitem{MUB_3}A. O. Pittenger and M. H. Rubin,
\newblock Linear Algebra Appl. \textbf{390}, 255 (2004).
%, \textit{Mutually Unbiased Bases, Generalized Spin Matrices and Separability},

\bibitem{MUB_4}
T. Durt, B.-G. Englert, I. Bengtsson, and K. Życzkowski,
\newblock Int. J. Quantum Inform. \textbf{08}, 535 (2010).
%ON MUTUALLY UNBIASED BASES

\bibitem{MUM_1}
A. Kalev and G. Gour, 
\newblock New J. Phys. \textbf{16}, 053038 (2014).
%Mutually Unbiased Measurements in Finite Dimensions

\bibitem{MUM_2}
M. Salehi, S. J. Akhtarshenas, M. Sarbishaei, and H. Jaghouri,
\newblock Quantum Inf. Process. \textbf{20}, 401 (2021).
%, \textit{Mutually Unbiased Measurements with Arbitrary Purity}

\bibitem{MUM_3}
M. Farkas, J. Kaniewski, and A. Nayak,
\newblock IEEE Trans. Inform. Theory \textbf{69}, 3814 (2023).
%, \textit{Mutually Unbiased Measurements, Hadamard Matrices, and Superdense Coding}


\bibitem{SIC_1}
J. M. Renes, R. Blume-Kohout, A. J. Scott, and C. M. Caves,
\newblock J. Math. Phys. \textbf{45}, 2171 (2004).
%\textit{Symmetric Informationally Complete Quantum Measurements}

\bibitem{SIC_2}
A. J. Scott and M. Grassl, 
\newblock J. Math. Phys. \textbf{51}, 042203 (2010).
%, Symmetric Informationally Complete Positive-Operator-Valued Measures: A New Computer Study

\bibitem{SIC_3}
G. Gour and A. Kalev,
\newblock J. Phys. A: Math. Theor. \textbf{47}, 335302 (2014).
%, Construction of All General Symmetric Informationally Complete Measurements


%----measurements used in entanglement detection---

\bibitem{ED_source_and_quantity}
C. Spengler, M. Huber, S. Brierley, T. Adaktylos, and B. C. Hiesmayr, 
\newblock Phys. Rev. A \textbf{ 86}, 022311 (2012).
%Entanglement detection via mutually unbiased bases

\bibitem{ED_use}A. E. Rastegin,
\newblock Open Syst. Inf. Dyn. \textbf{ 22}, 1550005 (2015). 
%On Uncertainty Relations and Entanglement Detection with Mutually Unbiased Measurements

\bibitem{huang_new}
S. Huang, W.-B. Liu, Y. Zhao, H.-L. Yin, Z.-B. Chen, and S. Wu,
\newblock arXiv:2210.00958.


\bibitem{ED_SIC}
Y. Xi, Z.-J. Zheng, and C.-J. Zhu, 
\newblock Quantum Inf Process \textbf{ 15}, 5119 (2016).
% Entanglement Detection via General SIC-POVMs
%-----------------------------------------------------------------------------------

\bibitem{MUB_tom_1}
W. K. Wootters, B. D. Fields,  
\newblock Ann. Phys. \textbf{ 191}, 363–381 (1989). 
%Optimal state-determination by mutually unbiased measurements.

\bibitem{MUB_tom_2}
R. B. A. Adamson, A. M. Steinberg,  
\newblock Phys. Rev. Lett. \textbf{ 105}, 030406 (2010).
%Improving quantum state estimation with mutually unbiased bases.


\bibitem{SIC_tom}
C. M. Caves, C. A. Fuchs, R. Schack,  
\newblock J. Math. Phys. \textbf{ 43}, 4537–4559 (2002).
%Unknown quantum states: The quantum de Finetti representation.


\bibitem{MUB_QKD_1}
N. J. Cerf, M. Bourennane, A. Karlsson, N. Gisin, 
\newblock Phys. Rev. Lett. \textbf{ 88}, 127902 (2002). 
% Security of quantum key distribution using d-level systems.

\bibitem{MUB_QKD_2}
I.-C. Yu, F.-L. Lin, C.-Y. Huang,  
\newblock Phys. Rev. A \textbf{ 78}, 012344 (2008). 
%Quantum secret sharing with multilevel mutually (un) biased bases.

\bibitem{SIC_QKD}
F. Bouchard, K. Heshami, D. England, R. Fickler, R. W. Boyd, B.-G. Englert, L. L. Sánchez-Soto, E. Karimi,  \newblock Quantum \textbf{ 2}, 111 (2018).
%Experimental investigation of high-dimensional quantum key distribution protocols with twisted photons.




\bibitem{heisenberg}
W. Heisenberg,
\newblock Z. Phys. \textbf{ 43}, 172 (1927).

	


\bibitem{EUR_1}
I. Białynicki-Birula and J. Mycielski,
\newblock  Commun.Math. Phys. \textbf{44}, 129 (1975).


\bibitem{EUR_2}
D. Deutsch,
\newblock Phys. Rev. Lett.  \textbf{50}, 631 (1983).


\bibitem{EUR_notable}
H. Maassen and J. B. M. Uffink,
\newblock Phys. Rev. Lett.  \textbf{60}, 1103 (1988).


\bibitem{shannon}
C. E. Shannon, 
\newblock Bell Syst. Tech. J. \textbf{ 27}, 379 (1948).


%next paragraph

\bibitem{EUR_review_1}
S. Wehner and A. Winter,
\newblock New J. Phys. \textbf{ 12}, 025009 (2010).

\bibitem{EUR_review_2}
P. J. Coles, M. Berta, M. Tomamichel, and S. Wehner,
\newblock Rev. Mod. Phys. \textbf{ 89}, 015002 (2017).

\bibitem{quantum_cryptography_review}
N. Gisin, G. Ribordy, W. Tittel, and H. Zbinden,
\newblock Rev. Mod. Phys. \textbf{ 74}, 145 (2002).


\bibitem{quantum_cryptography_1}
R. K\"{o}nig, S. Wehner, and J. Wullschleger,
\newblock IEEE Trans. Inf. Theory \textbf{ 58}, 1962 (2012).

\bibitem{quantum_cryptography_2}
F. Dupuis, O. Fawzi, and S. Wehner,
\newblock IEEE Trans. Inf. Theory \textbf{ 61}, 1093 (2015).





%---------------------not IC-------------------

\bibitem{ref_43}
U. Larsen, 
\newblock J. Phys. A: Math. Gen. \textbf{23}, 1041 (1990). 
%, Superspace geometry: the exact uncertainty relationship between complementary aspects

\bibitem{multiple_1}
I. D. Ivanovic, 
\newblock J. Phys. A: Math. Gen. \textbf{25}, L363 (1992).
%, \textit{An Inequality for the Sum of Entropies of Unbiased Quantum Measurements}



\bibitem{multiple_5}
M. A. Ballester and S. Wehner, 
\newblock Phys. Rev. A \textbf{75}, 022319 (2007).
%, \textit{Entropic Uncertainty Relations and Locking: Tight Bounds for Mutually Unbiased Bases}


\bibitem{multiple_6}
S. Liu, L.-Z. Mu, and H. Fan, 
\newblock Phys. Rev. A \textbf{91}, 042133 (2015). 
%Entropic uncertainty relations for multiple measurements


\bibitem{multiple_7}
B.-F. Xie, F. Ming, D. Wang, L. Ye, and J.-L. Chen,
\newblock Phys. Rev. A \textbf{104}, 062204 (2021).
%Optimized entropic uncertainty relations for multiple measurements,




%------------------------- IC-------------------

\bibitem{multiple_3}
 J. Sánchez-Ruiz,
 \newblock Phys. Lett. A \textbf{201}, 125 (1995). 
 %, Improved bounds in the entropic uncertainty and certainty relations for complementary observables


\bibitem{wu}
S. Wu, S. Yu, and K. M{\o}lmer,
\newblock Phys. Rev. A \textbf{ 79}, 022104 (2009).


\bibitem{MUB_SIC}
A. E. Rastegin,
\newblock Eur. Phys. J. D \textbf{ 67}, 269(2013).

\bibitem{ref_48}
B. Chen and S.-M. Fei, 
\newblock Quantum Inf. Proc. \textbf{14}, 2227 (2015). 
%, Uncertainty relations based on mutually unbiased measurements


\bibitem{huang_old}
S. Huang, Z.-B. Chen, and S. Wu, 
\newblock Phys. Rev. A \textbf{ 103}, 042205 (2021).

\bibitem{ref_42}
A. E. Rastegin,
\newblock Proc. R. Soc. A \textbf{479}, 20220546 (2023).
%, Entropic uncertainty relations from equiangular tight frames and their applications

\bibitem{huang_new_}
S. Huang, H.-L. Yin, Z.-B. Chen, and S. Wu, 
\newblock arXiv:2309.16955.
%\textit{Entropic Uncertainty Relations for Multiple Measurements Assigned with Biased Weights}, 

%----for quantum designs----

\bibitem{ket}
A. Ketterer and O. Gühne,
\newblock Phys. Rev. Research \textbf{ 2}, 023130 (2020).

\bibitem{ras}
A. E. Rastegin,
\newblock J. Phys. A: Math. Theor. \textbf{ 53}, 405301 (2020). 






\bibitem{renyi}
A. Rényi, 
\newblock \textit{Proceedings of the 4th Berkeley Symposium on Mathematical Statistics and Probability} (University of California Press, Berkeley, CA, 1961), Vol. 1, pp. 547.






%introduction is over



\bibitem{harr}
A. Haar, 
\newblock Ann. Math. \textbf{ 34}, 147 (1933).
%Der Massbegriff in Der Theorie Der Kontinuierlichen Gruppen




%---- introduce entanglement
%entanglement_cryptography
\bibitem{entangle_cryp_1}
A. K. Ekert,
\newblock Phys. Rev. Lett. \textbf{67}, 661 (1991).
%, \textit{Quantum Cryptography Based on Bell’s Theorem}

\bibitem{entangle_cryp_2}
C. H. Bennett, G. Brassard, and N. D. Mermin,
\newblock Phys. Rev. Lett. \textbf{68}, 557 (1992).
%, \textit{Quantum Cryptography without Bell’s Theorem}

\bibitem{entangle_cryp_3}
H.-K. Lo, M. Curty, and B. Qi,
\newblock Phys. Rev. Lett. \textbf{108}, 130503 (2012).
%, \textit{Measurement-Device-Independent Quantum Key Distribution}

%entanglement_random
\bibitem{entangle_rand}
S. Pironio et al.,
\newblock Nature \textbf{464}, 1021 (2010).
%, \textit{Random Numbers Certified by Bell’s Theorem}



\bibitem{tsallis}
C. Tsallis,
\newblock J. Stat. Phys. \textbf{52}, 479 (1988).

\bibitem{entanglement}
O. Gühne and M. Lewenstein, 
\newblock Phys. Rev. A \textbf{70}, 022316 (2004).
%, \textit{Entropic Uncertainty Relations and Entanglement}


\bibitem{steering}
T. Kriváchy, F. Fröwis, and N. Brunner,
\newblock Phys. Rev. A \textbf{98}, 062111 (2018).
%, \textit{Tight Steering Inequalities from Generalized Entropic Uncertainty Relations}


\bibitem{unbiasedness}
I. Bengtsson,
\newblock AIP Conf. Proc. \textbf{ 889}, 40(2007).


\bibitem{isomorphism}M.-D. Choi,
\newblock Linear Algebra Appl. \textbf{ 10}, 285 (1975).




\end{thebibliography}
\end{document}